\documentclass[aps,pra,preprint,superscriptaddress]{revtex4}

\usepackage{graphicx}
\usepackage{amssymb}
\newcommand {\eps}{\epsilon}

\small
\begin{document}

\title{Van der Waals  torque induced by external magnetic fields}

\author{R. Esquivel-Sirvent}
\email[Corresponding author. Email:]{raul@fisica.unam.mx}
\affiliation{Instituto de F\'{\i}sica, Universidad Nacional Aut\'onoma
 de M\'exico, Apartado Postal 20-364, D.F. 01000,  M\'exico}

\author{G. H. Cocoletzi}
\affiliation{Instituto de F\'{i}sica, Universidad Aut\'onoma de Puebla, Apartado  Postal J-48, Puebla 72570, M\'exico.}

 \author{M. Palomino-Ovando}
 \affiliation{Facultad de Ciencias Fisico-Matem\'aticas, Universidad Aut\'onoma de Puebla, Apartado Postal 5214, Puebla 72000, M\'exico.}
\date{\today}

\begin{abstract}

We present a method for inducing and controlling van der Waals torques between two parallel slabs using a constant magnetic field.  The torque is calculated using the Barash theory of dispersive torques. 
In  III-IV semiconductors such as $InSb$,  the effect of an external magnetic field is to induce an optical anisotropy, in an otherwise isotropic material,  that will in turn induce a torque.
 The calculations of the torque are done in the Voigt configuration, with 
the magnetic field  parallel to the surface of the slabs.  As a case  study we consider a slab made of calcite and a second slab made of $InSb$.  In the absence of magnetic field there is no torque. As the magnetic field increases, the optical anisotropy of $InSb$ increases and the torque becomes different from zero, increasing with the magnetic field.  The resulting torque  is  of the same order of magnitude as that calculated using permanent anisotropic materials when the  magnetic fields is close to 1 T.

\end{abstract}

\maketitle

\section{Introduction}

The attraction between two parallel neutral perfectly conducting plates separated by a gap was first predicted by Casimir in 1948 \cite{Casimir48}. This force was explained in terms of quantum vacuum fluctuations and is a well known result in quantum field theory \cite{Greiner}.  For two plates made of an arbitrary material Lifshitz \cite{Lifshitz56,Lifshitz61} derived a more general theory for dispersive forces based on Rytov's formalism for fluctuating electromagnetic fields \cite{Rytov67,Vinogradov09}. Solving Maxwell's equations with the proper boundary conditions, and using the fluctuation dissipation theorem, the force can be written in terms of the reflection coefficients between the gap and the slabs.
   Lifshitz theory is applicable at any temperature and separation between the bodies, with the restriction that the separation between the bodies be larger than the atomic separation.  For short separations the Lifshitz formula gives the non retarded Van der Waals force.  At  large separations retardation is included in the theory giving the Casimir force.
   The success of the Lifshitz formula is evident from the variety of systems that have been studied ranging from ferromagnets\cite{Bruno02}, semiconductors \cite{Lamoreaux08,Dalvit09} and metamaterials \cite{Rosa08,Footnote1}. 

Renewed interest in the  Casimir force came from  measurements  made in the mid 90's.  Lamoreaux \cite{Lamoreaux97} using a  torsional balance and Mohideen \cite{Mohideen98} with an atomic force microscope did the first precise measurements of the Casimir force. Later the use of micro torsional balances was implemented by several authors \cite{Dec03,Ian04,Vanzwol09}. With the exception of one experiment \cite{Onofrio02}, all the measurements are between a large sphere and a plane rather than between two parallel plates.   Recently  possible measurements of the Casimir force between a plane and a cylinder have been considered \cite{Wei10}.

It has also been suggested that dispersive forces could play an important role in the operation of micro electro mechanical
systems (MEMS) \cite{Serry95,Serry98, Roukes01}.   Experiments show that Casimir forces have a strong influence on the oscillatory behavior of microstructures, driving them into their nonlinear modes \cite{Capasso01}.  The study of the  of Casimir and Van der Waals forces in MEMS, in particular its role in pull-in dynamics has been also a topic of intensive study \cite{Zhao03,Zhao07,Delrio05,Esquivelapl,Esquivelnjp,Batra07}.   A possibility that has been considered by the authors \cite{Garcia09,Esquivel09,Esquivel10} is the use of external magnetic fields to decrease the magnitude of the Casimir force, and thus  inhibiting the pull-in or jump to contact.  

Lifshitz theory was developed for isotropic media. When anisotropic media are considered, an additional effect was shown to occur, there is a torque associated with the dispersive forces. In 1972 Parsegian and Weiss \cite{Parsegian72} derived an expression for the Van der Waals energy between anisotropic slabs in the general case when the medium between the slabs was also anisotropic. The resulting free energy depends on the angle between the principal axes of the slabs resulting in a torque.  Later, Barash \cite{Barash78} extended these results to include retardation effects predicting the Casimir torque. In the non retarded limit the results of Parsegian and Weiss were recovered.

 At the molecular level, the influence of  structural chirality on Van der Waals forces between two solids has been  studied by introducing an helical space variation of the dielectric tensor \cite{Galatry82}. This work was motivated by the need to understand the macroscopic properties of optically active species.

Different techniques have been used to derive the expression for the Van der Waals torque. For example, Zhoa \cite{Zhao05} calculated the Casimir torque using the method of quantized surface modes techniques, extending their results to anisotropic metamaterials \cite{Deng08}.  Philbin and Leonhardt \cite{Philbin08} calculated the electromagnetic stress tensor between anisotropic plates to obtain the same expression for the Casimir torque.
The Casimir torque has also been  derived as a consequence of angular momentum transfer of the fluctuating electromagnetic fields  between the two anisotropic planes \cite{Enk95,Torres06}.

Although dispersive torques have not been measured, Munday $et$ $al.$ \cite{Munday05} made detailed calculations of the Casimir torque, and proposed an experiment to measured the torque between a barium titanate disk immersed in ethanol  and a second  anisotropic material.   By means of an external laser the top plate is initially rotated letting the Casimir torque realigning the optical axes of the plates again. This paper showed that within current experimental techniques, dispersive torques can be measured.

In this paper we show that a dispersive torque can be induced using a constant  magnetic field, and the magnitude of the torque depends on the intensity of the magnetic field.  This is achieved by the excitation of magnetoplasmons in II-IV semiconductors such as $InSb$.

\section{Anisotropy due to external magnetic fields}

When a dc external magnetic field is applied to a metal or a highly doped semiconductor, the normal modes of the free charge change significantly giving rise to magneto plasmons. For example, the external magnetic field induces an optical anisotropy in an otherwise isotropic material. Also, the dispersion relation of the magneto plasmons presents a frequency gap that depends on the intensity of the external magnetic field \cite{Palik70,Wames72,Wallis74,Aers78}. Typically, magnetoplasmons are excited in III-V semiconductors such as InAb, InAs or GaAs.   Magnetoplasmons have reemerge in the context of plasmonics and its applications \cite{Kong08,Berman08,Liu09}.

To illustrate the effect of the external magnetic field, consider the classical equation of motion for the electrons in the material:
\begin{equation}
m\frac{d{\bf v}}{dt}=q({\bf E}+{\bf v}\times {\bf B})-\frac{m}{\tau}{\bf v}
\label{drude}
\end{equation}
where $m$ is the effective mass of the electron, $q$ the charge and $\tau$ is the relaxation time.  Assuming an harmonic electric field $e^{-i\omega t}$ the current ${\bf j}=nq{\bf v}$ can be found and thus the conductivity can be calculated \cite{Palik70}
\begin{equation}
\sigma_{ij}(\omega,{\bf B}_0)=\frac{nq^2}{\tau^* m}\frac{\delta_{ij}+\omega_c\tau^* \textit{e}_{ijk}(B_k/B_0)+(wc\tau^*)^2 (B_iB_j/B_0^2)}{1+(\omega_c \tau^*)^2},
\end{equation}
where $\tau^*=\tau/(1-i\omega \tau)$,  $w_c=q|{\bf B}_0|/mc$ is the cyclotron frequency, and $\textit{e}_{ijk}$ is the Levi-Civita symbol.
The dielectric tensor is obtained from
\begin{equation}
\epsilon_{ij}(\omega,{\bf B}_0)=\delta_{ij}+\frac{4 \pi i}{\omega}\sigma_{ij}.
\label{epsilon}
\end{equation}
Clearly if ${\bf B}_0=0$ we recover the results for the isotropic case.

For an arbitrary direction of the magnetic field, the calculation of the dispersion relation of the surface magneto plasmons and of the  optical reflectivity is difficult. To simplify the problem, specific directions of the magnetic field have to be chosen \cite{Manvir06}. In the so called {\it Faraday} configuration, the magnetic field is perpendicular to the slab. In this case,  there is mode conversion upon reflection from the slab. That is, for an incident  TE  wave, the reflected wave will consist of TE and TM modes, similar for an incident TM mode.

The second configuration, that will be used in this paper,  is the {\it Voigt} geometry where the magnetic field is parallel to the slabs. In this case there is no mode conversion upon reflection.   Consider a slab parallel to the $x-z$ plane. In the Voigt geometry the external magnetic field points along the $z$ axis.  In this case, the components of the dielectric tensor are given by \cite{Manvir06,Garcia09}
\begin{eqnarray}
\epsilon_{xx}&=&\epsilon_L\left[ 1-\frac{\omega_p^2}{\omega(\omega+i \gamma)} \right ], \nonumber \\
\epsilon_{yy}&=&\epsilon_L\left[ 1-\frac{(\omega+i\gamma)\omega_p^2}{\omega(\omega+i\gamma)^2-\omega_c^2} \right ] ,\nonumber \\
\epsilon_{yz}&=&\epsilon_L\left[ \frac{i\omega_c\omega_p^2}{\omega ((\omega+i\gamma)^2-\omega_c^2)} \right ],
\end{eqnarray}
and  $\epsilon_{zz}=\epsilon_{yy}$ and $\epsilon_{zy}=-\epsilon_{yz}$.  The other components are equal to zero.
In these equations $\epsilon_L$ is the background dielectric function, $\omega_p$ the plasma frequency and $\gamma$ is the damping parameter.  The factor $\epsilon_L$ accounts for the fact that we are working with semiconductors. 
In the absence of the magnetic field, $\omega_c=0$ and the plates become isotropic.  In the rest of the paper we will use the dimensionless variable $\Omega_c=\omega_c/\omega_p$, that  gives the relative importance of the external magnetic field. In Figure (2) we have plotted the dielectric function  components as given by  Eq. (4),   for a value of $\Omega_c=0.2$, showing the anisotropy of the system.   The parameters used are $\epsilon_L=15.8$,  $m=0.014 m_0$ and $\gamma=\omega_p/100$, that  are typical of $InSb$ \cite{Cunningham74}. In this figure we performed a  rotation of the frequency to the complex plane $\omega\rightarrow i\zeta$,  a common and convenient practice when calculating dispersive forces.  For clarity, the relation between the values of the parameter $\Omega_c$ and the magnetic field for $InSb$ are presented in the Appendix.

\section{Torque of the Van der Waals force}

In this section we briefly review the theory for calculating the torque associated with the Casimir force.  We follow the formalism of Barash \cite{Barash78}.
Let us consider two anisotropic plates labeled $i=1,2$, as depicted in Figure (1). Their optical axes are not aligned and form an  angle $\theta$.  The dielectric tensors of each plate can be described by

\begin{equation}
\left( \begin{array}{ccc}
\epsilon_{1 ||} & 0 & 0 \\
0& \epsilon_{1 \bot} & 0 \\
0& 0 & \epsilon_{1 \bot} \end{array} \right)
\end{equation}

and

\begin{equation}
\left( \begin{array}{ccc}
\epsilon_{2 ||} cos^2(\theta)+\epsilon_{2 \bot} sin^2(\theta)& (\epsilon_{2\bot}-\epsilon_{2 ||})sin(\theta)cos(\theta) & 0 \\
(\epsilon_{2\bot}-\epsilon_{2 ||})sin(\theta)cos(\theta)& \epsilon_{2 ||} sin^2(\theta)+\epsilon_{2 \bot} cos^2(\theta)& 0 \\
0& 0 & \epsilon_{2 \bot} \end{array} \right)
\end{equation}

In the case of anisotropic plates the free energy of the system depends on the separation between the plates and the angle between their optical axes,  this is ${\cal F}(\theta,L)$. The usual magnitude of the attractive force per unit area  is obtained from $F(\theta,L)=-\partial {\cal F}(\theta,L)/\partial L$, and the torque from $\tau(\theta,L)=-\partial {\cal F}(\theta,L)/\partial \theta$.  Although the calculation of the free energy for anisotropic systems is complicated, some simplifications are possible, depending on the separation between the plates and the degree of anisotropy.
The degree of anisotropy is quantified using the relation

\begin{equation}
\delta=\left |\frac{\epsilon_{||}}{\epsilon_{\bot}}-1\right |.
\label{delta}
\end{equation}
For $\delta<1$ we are in the low anisotropic regime. In Figure (2), we plot the values of $\delta$ for different values of the parameter $\Omega_c$. The values of $\Omega_c$ chosen also correspond to magnetic fields attainable in current laboratories.
In all cases, the anisotropy is small and an approximate expression for the torque is possible in the non retarded regime, and is given by \cite{Barash78,Munday05}
\begin{equation}
\tau(L,\theta)=-\frac{\hbar S}{64 \pi^2 L^2} \bar{w} sin(2\theta),
\label{torque}
\end{equation}
where $S$ is the surface of the plates and $\bar{w}$ is given by \cite{Munday05}.

\begin{equation}\label{barw}
\bar{w}=\int_0^{\infty} d\zeta \frac{(\eps_{2 ||}-\eps_{2\bot})(\eps_{1||}-\eps_{1 \bot}) \eps_3^2}{(\eps_{1 \bot}^2-\eps_3^2)(\eps_{2\bot}^2-\eps_3^2)}\times ln \left ( 1-\frac{(\eps_{1 \bot}-\eps_3)(\eps_{2\bot}-\eps_3)}{(\eps_{1 \bot}+\eps_3)(\eps_{2\bot}+\eps_3)}\right ).
\end{equation}

This expression considers that between the plates there could be a fluid with a dielectric function $\epsilon_3$ and that  temperature effects are not important and the separation $L$ satisfies $L<<c \hbar/T$.

Consider  the system depicted in Figure (1).  Let us assume a fixed bottom plate made of $InSb$ and the top plate made of a permanent anisotropic material such as calcite.  The space between both plates is ethanol. Both the calcite,  and the liquid between the plates were chosen to compare the effect of an external magnetic field with the torque calculations of Ref. (\cite{Munday05}). The  dielectric functions of calcite and ethanol are calculated using a two oscillator model with the main absorption contributions coming from the ultraviolet and infrared. The parameters used are those of Ref. (\cite{Munday05}).
Using Eqs. (\ref{torque},\ref{barw}) we calculate the torque for different values of $\Omega_c$ keeping the plate separation fixed at $L=100$ $ nm$.  In Figure (4) we show the torque per unit area ($\tau/s$). As the value of $\Omega_c$ increases, does so the torque.  Increasing $\Omega_c$ can be achieved by increasing the magnetic field, or by keeping the magnetic field constant and decreasing the number of carriers (see Appendix ).
The dependence of the torque on the applied magnetic field can be also seen in Figure (5). In this figure, we present the torque per unit area as a function of $\Omega_c$ keeping the angle between the plates fixed at $\theta=\pi/4$. The top curve is for a separation between the plates of $50$ $nm$ and the bottom curve for a separation of $100$ $nm$.

\section{Conclusions}
We have presented calculations of the Casimir torque, between two parallel plates, induced by an external magnetic fields in the Voigt configuration. As a case  study, we calculated the torque between calcite and $InSb$ with the space between the slabs filled with ethanol. At zero magnetic field there is no torque since the $InSb$ slab is isotropic. Applying an external magnetic field induces an optical anisotropy, inducing a torque.  This  system allows a direct comparison with previous calculations and suggested experiments  of Ref.(\cite{Munday05}), thus showing the viability also of measuring the Van der Waals torque induced by an external magnetic field.  The magnetic fields needed to induce the torque are easily obtained in current laboratories. The largest magnetic field used in our calculations was of 2.1 T.

\section{ACKOWLEDGEMENTS} The work of GHC was supported partially by VIEP-BUAP Project DES-EXC.  Partial support from DGAPA-UNAM project no. IN 113208,  CONACyT project no.  82474, and from the NERC EFRC of the US DOE (BES Award DE-SC0000989).  We thank Prof. George Schatz from Northwestern University for his hospitality during the conclusion of this paper and for his careful reading and criticisms of the manuscript.

\section{Appendix}

The introduction of the dimensionless frequency $\Omega_c$ has the advantage that its value can be changed by changing either the magnetic field or the number density of the electrons in the material. From the definition of
the plasma frequency and the cyclotron frequency,  we have

\begin{equation}
\label{omegaca}
\Omega_c=\frac{|{\bf B}_0|}{c \sqrt{4\pi n m}}.
\end{equation}

For $InSb$ with $m=0.014m_0$ and $n=2\times10^{16}$$cm^{-3}$, we have $\Omega_c\sim1.862\times10^{-5}  |{\bf B}_0|$.
The values used in Figure 2 for $\Omega_c=0.1,0.2,0.4$ correspond to the magnetic fields $|{\bf B}_0|\sim 0.5,1.0, 2.1$ $Teslas$, well within reach of common laboratory magnets.

From Eq. (\ref{omegaca})  we see that the value of $\Omega_c$ can also be changed  by keeping the magnetic field fixed and changing the number of charge carriers in the sample.

 \newpage

 \begin{figure}[h]
\includegraphics[width=.7\textwidth]{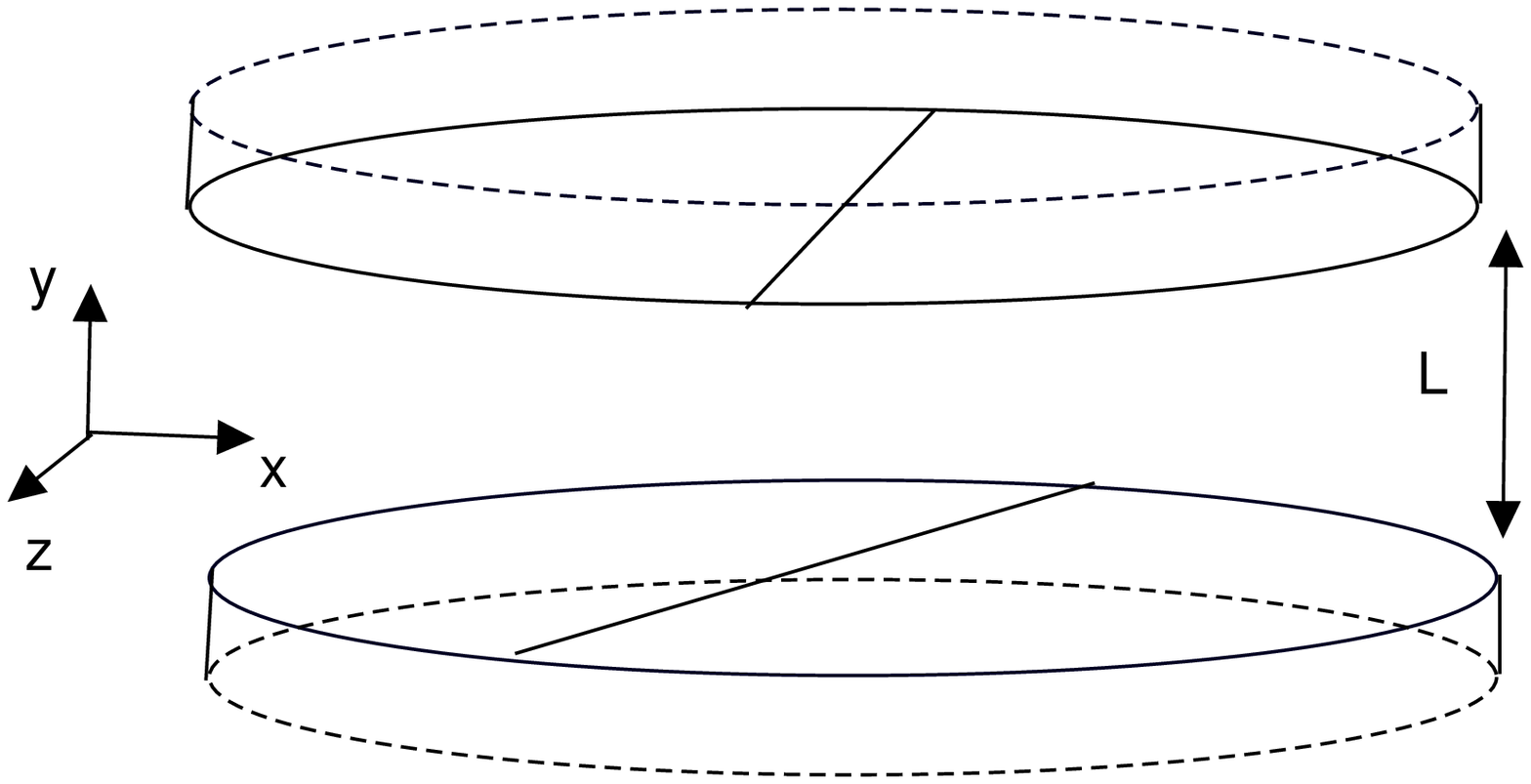}
\label{fig1}
\caption{ Parallel plate configuration. For anisotropic systems, the principal axes of the  plates make an   angle $\theta$.  }
\end{figure}

 \begin{figure}[h]
\includegraphics[width=.7\textwidth]{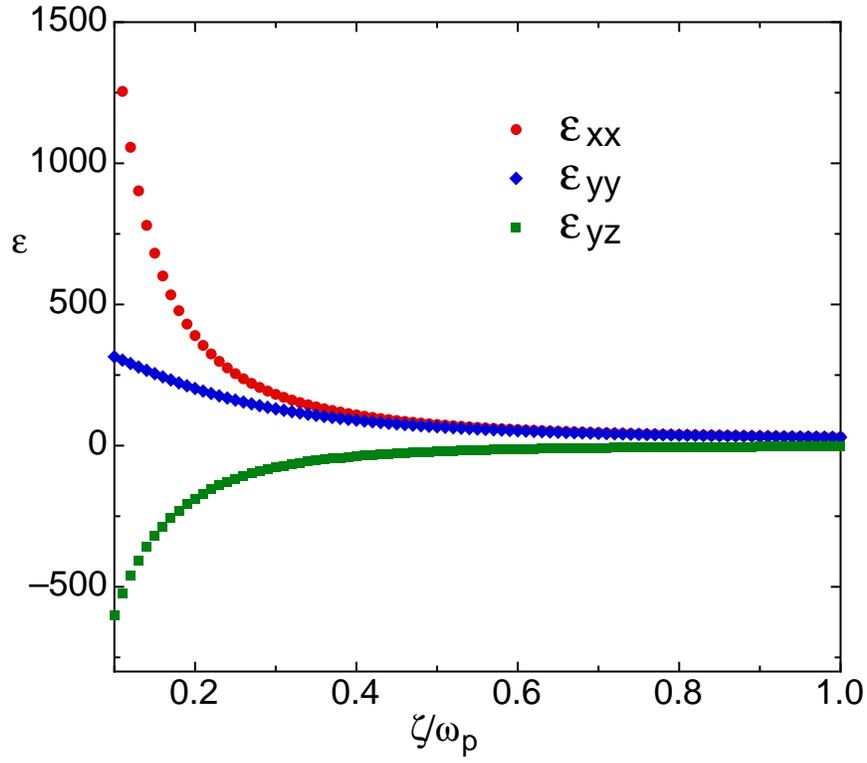}
\label{fig2}
\caption{ Components of the dielectric tensor as a function of the imaginary frequency $\zeta$, normalized to the plasma frequency $\omega_p$. The components of the dielectric tensor correspond to $InSb$ and for $\Omega_c=0.2$.  }
\end{figure}

 \begin{figure}[h]
\includegraphics[width=.7\textwidth]{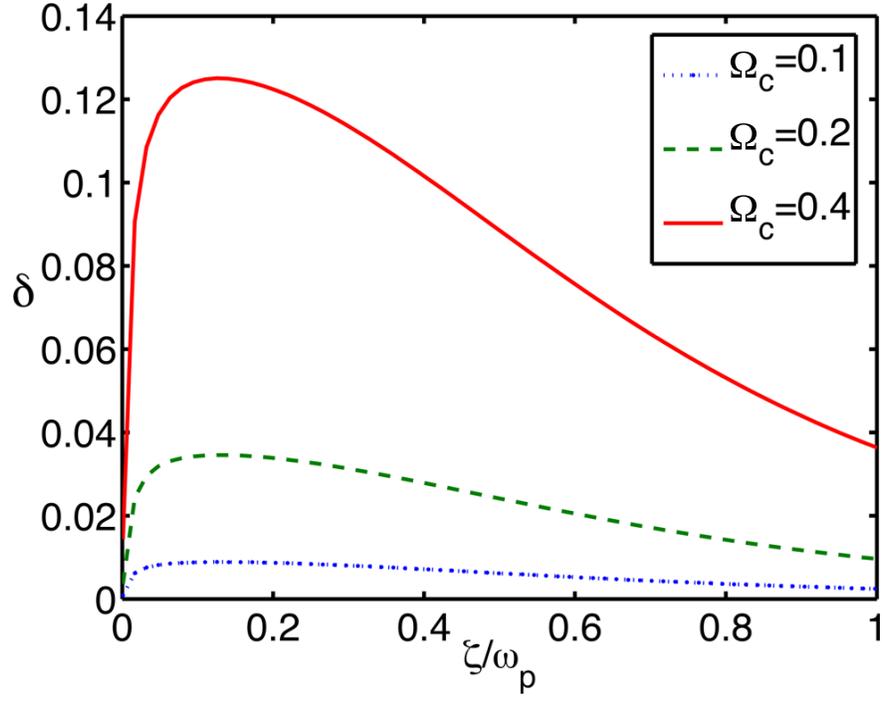}
\label{fig3}
\caption{ Degree of anisotropy $\delta$, Eq. (\ref{delta}) of $InSb$ for different values of the reduced frequency $\Omega_c$.  With the chosen values, we see that the plate made of $InSb$ has a small anisotropy.  }
\end{figure}

\begin{figure}[h]
\includegraphics[width=.7\textwidth]{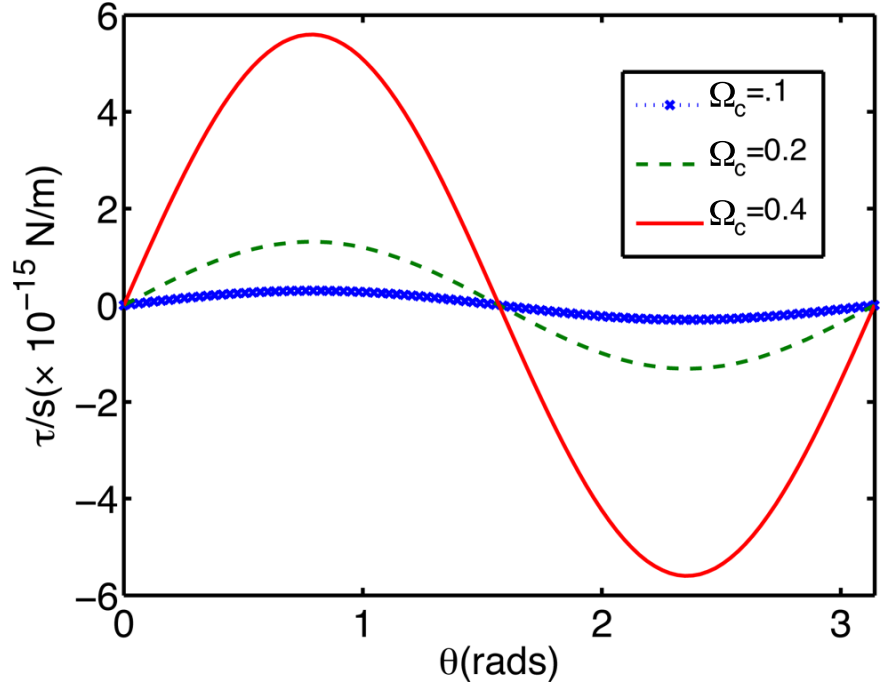}
\label{fig4}
\caption{Torque per unit area as a function of angle when the top plate is made of calcite, the bottom plate of $InSb$ and assuming that the space between the plates is filled with ethanol. The dielectric function for calcite and ethanol were taken from Ref. (\cite{Munday05}). Different curves correspond to different values of the reduced frequency $\Omega_c$. As the magnetic field increases, so does $\Omega_c$,  inducing a larger anisotropy and hence a larger torque. In these curves the separation between the plates was kept fixed at $d=100$ $nm$. }
\end{figure}

\begin{figure}[h]
\includegraphics[width=.7\textwidth]{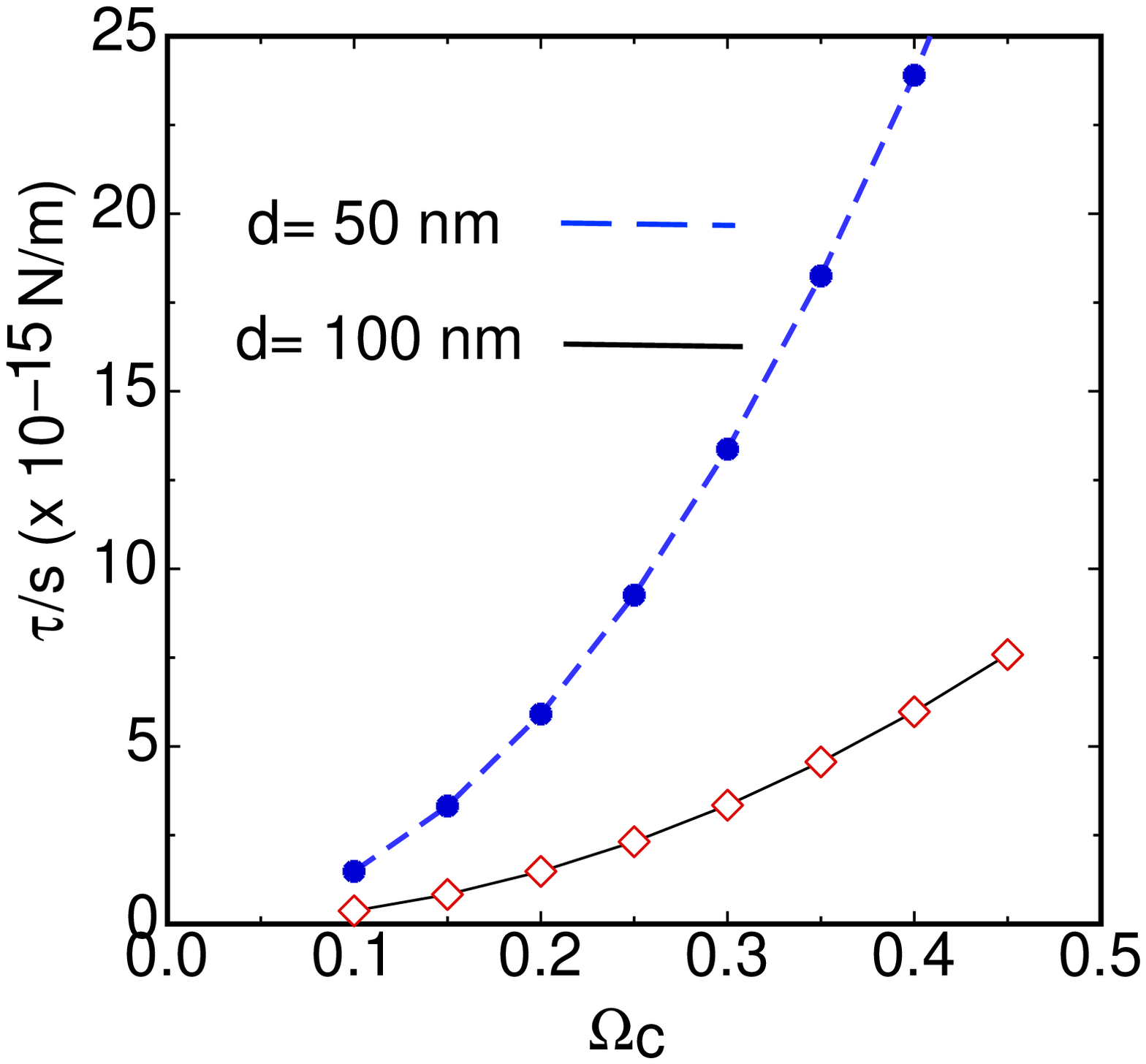}
\label{fig5}
\caption{Keeping the rotation angle fixed at $\theta=\pi/4$, the torque per unit area as a function of $\Omega_c$ is presented. The top curve correspond to a fixed separation between the plates of $L=50$ $nm$ and the bottom curve for a separation of $L=100$ $nm$. We see that decreasing the separation not only increases the force but also the torque.  }
\end{figure}


\begin{thebibliography}{99}
\bibitem{Footnote1} The papers so far cited are only a small sample of the vast amount of publications on the Casimir effect in the last few years.
\end{thebibliography}
\end{document}